\documentclass[preprint]{revtex4}
\usepackage{graphicx}

\begin{document}
\title{Constraint on the Squeeze Parameter of Inflaton from Cosmological Constant}

\author{Sang Pyo Kim}
\email{sangkim@kunsan.ac.kr}
\author{Seoktae Koh}
\email{kohst@ihanyang.ac.kr} \affiliation{Department of Physics,
Kunsan National University, Kunsan, 573-701,Korea}

\date{\today}

\begin{abstract}
The inflaton is highly likely to settle in a squeezed vacuum state
after inflation. The relic inflaton after inflation and reheating
undergoes a damped oscillatory motion and contributes to the
effective cosmological constant. We interpret the renormalized
energy density from the squeezed vacuum state as an
effective cosmological constant. Using the recent observational
data on the cosmological constant, we find the constraint on the squeeze parameter of the
inflaton in the early universe.
\end{abstract}

\maketitle

\section{Introduction}

Recent observations indicate that our present universe is
dominated by the dark energy with the pressure-to-density ratio
$p/\rho \lesssim -1/3$, and that our universe is nearly flat
($\Omega_0 \simeq 1$). One of the promising candidates for the dark
energy is the cosmological constant with $p=- \rho$. From the
field theory point of view, the vacuum energy density of fields
contributes to an effective cosmological constant and would have
been created through various phase transition processes  such as the inflation,
electro-weak, and QCD phase transitions. The effective vacuum energy
density, $\rho_V$, defined as \cite{weinberg89}
\begin{eqnarray}
\rho_V =\frac{\Lambda_{eff}}{8\pi G} = \langle \hat{\rho} \rangle
+ \frac{\Lambda}{8\pi G},
\end{eqnarray}
can be constrained by the present dark energy observation
\cite{wmap}
\begin{eqnarray}
\rho_V \leq \Omega_{DE} \rho_c,
\end{eqnarray}
where $\Omega_{DE}$ is the fraction of the dark energy density in
the present universe and $\rho_c$ is the present critical energy
density.

Inflation scenarios are a successful model of the early universe
that solves many puzzles of the standard cosmological model. As the
inflaton, a scalar field, rolled slowly over any nearly flat
potential, the universe would have undergone a phase of accelerated
expansion. The energy density of the inflaton during the slow-roll
over played the role of a cosmological constant. Any initial
vacuum state of the inflaton would have evolved to a squeezed
quantum state during an inflation period through the parametric
amplification \cite{grishchuk90}. When inflation ended, the
inflaton began to oscillate around the global minimum of its
effective potential and created particles and thus reheated the
universe. A more efficient mechanism for a cornucopia of particles
would be a preheating process where another scalar field couples
to the oscillating inflaton and undergoes a parametric resonance.
In either case, the inflaton evolved to a squeezed vacuum state.

Recently there has been an attempt to understand the cosmological
constant as a relic of vacuum fluctuations of the inflaton
\cite{choi04}. The energy density of the inflaton in a vacuum
state after the reheating process constitutes a significant fraction of
the cosmological constant. Even with the correct theory of
particle physics, there is still an ambiguity in choosing the vacuum
state. In this paper we study how a general
squeezed vacuum state affects the cosmological constant problem.
The renormalized energy density of the squeezed vacuum state that is obtained
by subtracting the vacuum energy density \cite{tanaka00} is interpreted as an effective cosmological constant. Further, we find a constraint on the squeeze parameter of the inflaton from the present observational data.

The organization of this paper is as follows. In Sec. II, we
employ a massive inflaton model to find the squeezed vacuum state
after the reheating period. The equation is found for the
evolution of the energy density of the squeezed vacuum state from the reheating period to the present epoch. In
Sec. III, using the energy density and the recent observational
data, we put the constraint on the squeeze parameter of the
inflaton. We compare our result with the squeeze parameter of
gravitational waves and trans-Planckian physics. In Sec. IV, we
discuss the physical implication of the squeeze parameter of the
inflaton.

\section{Squeezed Vacuum States of Inflaton}

We consider a simple inflation model which is described by a
single inflaton, a scalar field, and has the action
\begin{eqnarray}
S_{\phi}=-\int d^4 x \sqrt{-g} \left[\frac{1}{2}g^{\mu\nu}\partial_{\mu}\phi
\partial_{\nu}\phi+V(\phi)\right],
\label{action}
\end{eqnarray}
in the Friedmann-Robertson-Walker universe
\begin{eqnarray}
ds^2 = -dt^2 +a^2(t)dx^i dx_i.
\end{eqnarray}
After Fourier transforming the inflaton into the momentum space and
varying the action with respect to $\phi$, we get the equation of
motion for the ${\bf k}$-th mode
\begin{eqnarray}
\ddot{\phi}_{\bf k}+3 H \dot{\phi}_{\bf k}+\frac{k^2}{a^2}
\phi_{\bf k}+ V_{,\phi}=0,
\end{eqnarray}
where $H = \dot{a}/ a$ is a Hubble parameter, dots denote
derivatives with respect to $t$, and the subscript $\phi$ denotes
a derivative with respect to $\phi$. When the inflaton stops
slow-rolling over the potential, inflation ends and the inflaton
settles down and oscillates around the potential minimum. In this
regime the potential can be approximated up to quadratic order of
$\phi$ as
\begin{eqnarray}
V \simeq V_0 + \frac{1}{2}m^2 \phi^2,
\end{eqnarray}
where $m^2$ is the curvature of the potential with the
dimension of mass. Then the equation of motion becomes
\begin{eqnarray}
\ddot{\phi}_{\bf k}+3 H \dot{\phi}_{\bf k}+\omega_{\bf k}^2
\phi_{\bf k} =0, \quad \omega_{\bf k}^2 \equiv m^2+\frac{k^2}{a^2}.
\end{eqnarray}
The Hamiltonian density for the action (\ref{action}) is a
collection of time-dependent harmonic oscillators
\begin{eqnarray}
\hat{\mathcal{H}} = \sum_{\bf k}\hat{\mathcal{H}}_{\bf k}
      = \sum_{\bf k}\left[\frac{1}{2a^3}\hat{\pi}_{\bf k}^2
+\frac{a^3}{2} \omega_{\bf k}^2 \hat{\phi}_{\bf k}^2\right],
\label{hamil}
\end{eqnarray}
where $\pi_{\bf k} = a^3\dot{\phi}_{\bf k}$ is a conjugate momentum.

For the Hamiltonian density (\ref{hamil}), we may introduce the
time-dependent annihilation and creation operators (in units of
$\hbar = c = 1$) \cite{kim99,kim01,kim04}
\begin{eqnarray}
\hat{a}_{\bf k} (t) &=& i[\varphi^{\ast}_{\bf k}(t)\hat{\pi}_{\bf k}
-a^3\dot{\varphi}^{\ast}_{\bf k}(t)\hat{\phi}_{\bf k}], \nonumber \\
\hat{a}_{\bf k}^{\dag} (t) &=& - i[\varphi_{\bf k}(t) \hat{\pi}_{\bf
k} - a^3 \dot{\varphi}_{\bf k}(t)\hat{\phi}_{\bf k}],
\end{eqnarray}
where $\varphi_{\bf k}(t)$ is a complex solution to the classical
equation of motion
\begin{eqnarray}
\ddot{\varphi}_{\bf k} + 3 H\dot{\varphi}_{\bf k}+\omega_{\bf k}^2
\varphi_{\bf k} = 0. \label{eom}
\end{eqnarray}
These are invariant operators satisfying the quantum Liouville-von
Neumann equation
\begin{eqnarray}
i \frac{\partial}{\partial t}\hat{a}_{\bf k} +[\hat{a}_{\bf k},
\hat{\mathcal{H}}_{\bf k}] =0,
\end{eqnarray}
and so does $\hat{a}^{\dag}_{\bf k}$. Also these operators satisfy
the equal-time commutation relation
\begin{eqnarray}
[\hat{a}_{\bf k}, \hat{a}_{{\bf k}'}^{\dag}] = \delta_{{\bf k} {\bf k}'},
\end{eqnarray}
provided that the Wronskian condition
\begin{eqnarray}
a^3 (\varphi_{\bf k} \dot{\varphi}_{\bf k}^{\ast} - \varphi_{\bf
k}^{\ast} \dot{\varphi}_{\bf k}) =i
\end{eqnarray}
be imposed. Further, we may select a solution $\varphi_{\bf k} $ that
has the minimum uncertainty \cite{kim99}. Then the most general
complex solution to Eq. (\ref{eom}) is given by
\begin{eqnarray}
\varphi_{{\bf k}\nu}=\mu_{\bf k}^{\ast} \varphi_{\bf k} +
\nu_{\bf k}^{\ast}\varphi_{\bf k}^{\ast},
\end{eqnarray}
where
\begin{equation}
|\mu_{\bf k}|^2 - |\nu_{\bf k}|^2 =1. \label{bog rel}
\end{equation}
Here we set the parameters as
\begin{eqnarray}
\mu_{\bf k} = \cosh r_{\bf k}, \quad \nu_{\bf k} = e^{i\theta_{\bf
k}} \sinh r_{\bf k},
\end{eqnarray}
where $r_{\bf k}$ and $\theta_{\bf k}$ are two real squeeze
parameters for the ${\bf k}$-th mode.

The time-dependent annihilation and creation operators defined in
terms of $\varphi_{{\bf k} \nu}$ and $\varphi_{\bf k}$ are related
through the Bogoliubov transformation
\begin{eqnarray}
\hat{a}_{{\bf k} \nu} (t) &=& \mu_{\bf k} \hat{a}_{\bf k} (t) -\nu_{\bf
k} \hat{a}_{\bf k}^{\dag} (t)
=\hat{S}_{\bf k} (z_{\bf k}, t)\hat{a}_{\bf k} (t)
\hat{S}_{\bf k}^{\dag}(z_{\bf k}, t), \nonumber \\
\hat{a}_{{\bf k} \nu}^{\dag} (t) &=& \mu_{\bf k}^{\ast} \hat{a}_{\bf
k}^{\dag} (t) - \nu_{\bf k}^{\ast} \hat{a}_{\bf k} (t) =
\hat{S}_{\bf k} (z_{\bf k}, t)\hat{a}_{\bf k}^{\dag} (t) \hat{S}_{\bf
k}^{\dag}(z_{\bf k}, t),
\end{eqnarray}
where the squeeze operator $\hat{S}_{\bf k}(z_{\bf k}, t)$ is defined
by
\begin{eqnarray}
\hat{S}_{\bf k}(z_{\bf k}, t)=\exp\left[\frac{1}{2}(z_{\bf k}
\hat{a}_{\bf k}^{\dag 2} (t) -z_{\bf k}^{\ast} \hat{a}_{\bf
k}^2 (t))\right].
\end{eqnarray}
with $z_{\bf k} = r_{\bf k} e^{i\theta_{\bf k}}$. The squeezed
number state, which is an eigenstate of the squeezed number
operator, $\hat{a}_{{\bf k}\nu}^{\dag}\hat{a}_{{\bf k}\nu}$, is
defined by the squeeze number operator as
\begin{eqnarray}
|n_{\bf k},z_{\bf k}, t \rangle = \hat{S}_{\bf k}(z_{\bf
k}, t)|n_{\bf k}, t \rangle, \quad \hat{a}^{\dag}_{{\bf k} \nu} (t)
\hat{a}_{{\bf k} \nu} (t) |n_{\bf k}, z_{\bf k}, t \rangle =
n_{\bf k} |n_{\bf k}, z_{\bf k}, t \rangle.
\end{eqnarray}
Then the Hamiltonian density for the ${\bf k}$-th mode in Eq.
(\ref{hamil}) has the representation
\begin{eqnarray}
\hat{\mathcal{H}}_{\bf k} = \frac{a^3}{2}[ (\dot{\varphi}_{\bf
k}^2+\omega_{\bf k}^2 \varphi_{\bf k}^2)\hat{a}_{{\bf k} \nu}^2
+(\dot{\varphi}_{\bf k} \dot{\varphi}_{\bf k}^{\ast} +\omega_{\bf
k}^2 \varphi_{\bf k}\varphi_{\bf k}^{\ast}) (\hat{a}_{{\bf k} \nu}
\hat{a}_{{\bf k} \nu}^{\dag}+\hat{a}_{{\bf k}
\nu}^{\dag}\hat{a}_{{\bf k} \nu}) +(\dot{\varphi}_{\bf k}^{\ast
2}+\omega_{\bf k}^2 \varphi_{\bf k}^{\ast 2}) \hat{a}_{{\bf k}
\nu}^{\dag 2}],
\end{eqnarray}
from which follows the expectation value of the ${\bf k}$-th mode
energy density
\begin{eqnarray}
\langle \hat{\rho}_{\bf k} \rangle_n & = & \frac{1}{a^3} \langle
n_{\bf k}, z_{\bf k}, t |\hat{\mathcal{H}}_{\bf k}
|n_{\bf k}, z_{\bf k}, t \rangle \nonumber \\
&=& (n_{\bf k} + 1/2) \Bigl\{(|\mu_{\bf k}|^2+|\nu_{\bf k}|^2)
(\dot{\varphi}_{\bf k}\dot{\varphi}_{\bf k}^{\ast} +\omega_{\bf
k}^2 \varphi_{\bf k}\varphi_{\bf k}^{\ast}) + 2 {\mathcal Re}[
(\dot{\varphi}_{\bf k}^2+\omega_{\bf k}^2 \varphi_{\bf
k}^2)\mu_{\bf k}^{\ast}\nu_{\bf k}] \Bigr\}.
\end{eqnarray}
We can safely drop the second term of the right hand side in the
last line by employing, for instance, the random phase
approximation \cite{prokopec93}. The squeezed vacuum state result,
obtained by setting $n_{\bf k}=0$, is given by
\begin{eqnarray}
\langle \hat{\rho}_{\bf k} \rangle_{sv} = \frac{1}{2} (|\mu_{\bf
k}|^2+|\nu_{\bf k}|^2) (\dot{\varphi}_{\bf k}\dot{\varphi}_{\bf
k}^{\ast} +\omega_{\bf k}^2 \varphi_{\bf k}\varphi_{\bf
k}^{\ast}). \label{vev}
\end{eqnarray}
As the adiabatic vacuum state corresponds to $\mu_{\bf k} = 1$ and
$\nu_{\bf k} = 0$, the vacuum expectation value of the ${\bf
k}$-th mode is
\begin{eqnarray}
\langle \hat{\rho}_{\bf k} \rangle_{vac} = \frac{1}{2}
(\dot{\varphi}_{\bf k} \dot{\varphi}_{\bf k}^{\ast} +\omega_{\bf
k}^2 \varphi_{\bf k} \varphi_{\bf k}^{\ast}).
\end{eqnarray}

To calculate the $\langle \hat{\rho}_{\bf k} \rangle_{vac}$, we
need to solve the equation of motion (\ref{eom}). Changing the
variable
\begin{eqnarray}
\varphi_{\bf k} = a^{-3/2} u_{\bf k},
\end{eqnarray}
Eq. (\ref{eom}) can be written in a canonical form
\begin{eqnarray}
\ddot{u}_{\bf k} + \left(\omega_{\bf k}^2
-\frac{9}{4}H^2-\frac{3}{2}\dot{H} \right) u_{\bf k} =0.
\label{eom2}
\end{eqnarray}
After slow-rolling over the potential, the inflaton begins to
oscillate around the minimum and has an adiabatic solution of the form \cite{kim99b}
\begin{eqnarray}
u_{\bf k} (t) = \frac{1}{\sqrt{2\Omega_{\bf k} (t)}}e^{-i\int
\Omega_{\bf k}(t) dt},
\end{eqnarray}
where
\begin{eqnarray}
\Omega^2_{\bf k} = \left[\omega_{\bf k}^2
-\frac{9}{4}H^2-\frac{3}{2}\dot{H}
+\frac{3}{4}\frac{\dot{\Omega}_{\bf k}^2}{\Omega_{\bf k}^2}
-\frac{1}{2}\frac{\ddot{\Omega}_{\bf k}}{\Omega_{\bf k}}\right].
\end{eqnarray}
As in Ref. \cite{kim99b}, we shall assume $\omega_{\bf k} \gg H$,
which can be justified for the inflaton inside the horizon, and
also assume $\Omega_{\bf k} \gg |\dot{\Omega}_{\bf k}|,
|\ddot{\Omega}_{\bf k}|$ so that $\Omega_{\bf k} \approx
\omega_{\bf k}$. The vacuum energy density of the inflaton itself
is the sum of the vacuum energy of all the modes
\begin{equation}
\langle \hat{\rho} \rangle_{vac} = \sum_{\bf k} \langle
\hat{\rho}_{\bf k} \rangle_{vac} = \frac{1}{(2 \pi)^3} \int
\frac{\omega_{\bf k}}{2a^3} d^3 {\bf k}.
\end{equation}
To remove the ultraviolet divergences, we may introduce a cut-off
momentum $k_{\Lambda}$
\begin{eqnarray}
\langle \hat{\rho} \rangle_{vac} &=& \frac{4 \pi^2}{2(2 \pi)^3
a^3}
\int_{0}^{k_{\Lambda}} \sqrt{m^2 + \frac{k^2}{a^2}} dk \nonumber\\
&=& \frac{1}{16 \pi^2 a} \Biggl[ k_{\Lambda} \Bigl(
\frac{k_{\Lambda}^2}{a^2} + m^2 \Bigr)^{3/2} - \frac{m^2}{2}
k_{\Lambda} \Bigl( \frac{k_{\Lambda}^2}{a^2} + m^2 \Bigr)^{1/2}
\cdots \Biggr],
\end{eqnarray}
and regularize the infinite quantities with the cosmological
constant, gravitational constant, mass and {\it etc} \cite{kim97}. There is a simple method to get the renormalized value \cite{tanaka00}
\begin{equation}
\langle \hat{\rho} \rangle_{ren} = \langle \hat{\rho} \rangle_{sv}
- \langle \hat{\rho} \rangle_{vac} = \frac{1}{a^3} \sum_{\bf k}
|\nu_{\bf k}|^2 \omega_{\bf k}.
\end{equation}
Solving the semiclassical Einstein equation
\begin{eqnarray}
H^2 =\frac{8 \pi G}{3} \langle \hat{\rho} \rangle_{ren},
\end{eqnarray}
we get the leading term of the power-law expansion
\begin{eqnarray}
a(t) = \Biggl(3 \pi G_{ren}  \sum_{\bf k} |\nu_{\bf k}|^2
\omega_{\bf k} \Biggr)^{1/3}t^{2/3}. \label{sol}
\end{eqnarray}
where $G_{ren}$ denotes the renormalized gravitational constant.
This shows that as the inflaton settles down to the damped
oscillatory motion, the universe enters the matter-dominated era.
Using the power-law expansion (\ref{sol}), the squeezed vacuum
expectation value of the energy density evolves as
\begin{eqnarray}
\langle \hat{\rho} (t) \rangle_{ren} = \langle \hat{\rho} (t_r)
\rangle_{ren} \left(\frac{t_r}{t}\right)^2 \label{ev den}
\end{eqnarray}
where $\langle \hat{\rho} (t_r) \rangle_{ren}$ is the value at $t=t_r$.

\section{Constraint on the Squeeze Parameter from Observations}

Assuming inflation to occur at GUT scales, the energy and time
scales at the reheating time are approximately $ E_r \sim
10^{14}GeV$ and $t_r \simeq 10^{-34} sec$, respectively. Hence the
vacuum energy density at the reheating,
\begin{eqnarray}
\langle \hat{\rho}_{\bf k} (t_r) \rangle \geq (10^{14} GeV)^4,
\label{reh den}
\end{eqnarray}
evolves, according to Eq. (\ref{ev den}), to the present value
\begin{eqnarray}
\langle \hat{\rho} (t_0) \rangle_{ren} &\simeq& \sum_{\bf k}
|\nu_{\bf k}|^2 \omega_{\bf k} (t_r) \times
\left(\frac{10^{-34}}{4\times 10^{17}}\right)^2
\nonumber \\
&\gg& \sum_{\bf k}  |\nu_{\bf k}|^2 \times 6.25 \times 10^{-48}
(GeV)^4.
\end{eqnarray}
Here we used the age of the universe, $t_0 \simeq 4\times 10^{17}
sec$, from the recent observational data \cite{wmap}. In the last
line the inequality $\langle \hat{\rho}_{\bf k} (t_r) \rangle \gg
(10^{14} GeV)^4$ was used for high momentum mode. Again using the
data from observations
\begin{eqnarray}
\rho_{DE}(t_0) = \Omega_{DE} \rho_c \simeq 2.94 \times 10^{-47}
(GeV)^4
\end{eqnarray}
we put a constraint on the squeeze parameter
\begin{eqnarray}
\sum_{\bf k} |\nu_{\bf k}|^2 \ll  4.7. \label{cons1}
\end{eqnarray}

A few comments are in order. First, note that the constraint
(\ref{cons1}) on the squeeze parameter applies only to the period
after reheating and should be distinguished from another
constraint from the inflation period. Second, the number of
created pairs
\begin{eqnarray}
n_{\bf k} = |\nu_{\bf k}|^2 =\sinh^2 r_{\bf k}
\end{eqnarray}
is exponentially suppressed for high momentum modes when the
inflaton evolves adiabatically, that is, keeps the adiabatic
vacuum state throughout the evolution \cite{birrell82}. For instance,
the squeeze parameter of
gravitational waves (with $m = 0$)  which were produced from quantum
fluctuations and got amplified, was calculated by Grishchuk and
Sidorov during inflation period \cite{grishchuk90}. It was shown
that $r_{\bf k}$ goes to zero for wavelengths smaller than the
horizon size at the end of inflation but increases for wavelengths
larger the horizon size. The horizon-sized wavelength has $r_{\bf
k} \simeq 1$ for $f \simeq 10^8 Hz$ where $f$ corresponds to the
current frequency whereas the superhorizon-sized wavelength has $r_{\bf
k} \simeq 10^2$ for $f\simeq 10^{-16} Hz$. This means that the
high momentum modes with wavelength smaller than the horizon
remain in the adiabatic vacuum state whereas the superhorizon-sized
low momentum modes are highly squeezed due to the parametric
interaction with the gravitational field during inflation.

Finally, we compare our result with trans-Planckian physics. In
the very early universe before inflation, the inflaton belonged to
a ultra-relativistic regime and behaved like a massless field. A
large number of particles would have been created for wavelengths
smaller than the horizon size. Thus the squeeze parameter
$\nu_{\bf k}$ should be constrained to protect overproduction of
particle pairs at the end of the inflation
\cite{starobinsky01,tanaka00}
\begin{eqnarray}
|\nu_{\bf k}|^2 \ll  \Bigl(\frac{H}{M_{pl}} \Bigr)^2.
\label{cons2}
\end{eqnarray}
The present CMB data restricts to $H/M_{pl} < 10^{-5}$. Whereas
the constraint (\ref{cons1}) gives bounds for high momentum modes
\begin{eqnarray}
|\nu_{\bf k}|^2 \ll  \frac{C}{k^{3 + \epsilon}} \label{cons3}
\end{eqnarray}
for a positive $\epsilon$ and $C$ independent of the momentum.

\section{Summary and Discussion}

The inflaton provides a necessary energy not only to drive a
quasi-exponential expansion and but also to reheat the universe
during and after inflation. The inflaton would be properly
described by quantum theory (semiclassical gravity) in the early
universe. The initial vacuum state of the inflaton is dynamically
squeezed by the expansion of the universe and remains a squeezed
vacuum during and after reheating. Also it is likely that the
inflaton may survive as a relic after the reheating period and
undergo a damped oscillatory motion. Then the renormalized energy
density of the squeezed vacuum state contributes to the effective
cosmological constant and evolves according to the
matter-dominated power-law. The recent observational data on the
dark energy or cosmological constant may put a constraint on the
energy density at the reheating time. It is thus possible to find
the constraint on the squeeze parameter of the squeezed vacuum
state of the inflaton. It is important to know what quantum state
of the inflaton would be in the early universe. The initial
quantum state may affect on the temperature anisotropy
\cite{koh04}. This result of this paper may help tackle some of
discussions on the early universe.

\acknowledgements
The first author would like to thank Prof. C.I. Um
for useful discussions. This work was supported by Korea Research
Foundation under Grant No. 2000-015-DP0080.

\end{document}